\documentclass[%
 reprint,
 amsmath,amssymb,
 aps,
]{revtex4-2}

\usepackage{graphicx}
\usepackage{dcolumn}
\usepackage{bm}

\usepackage{amsmath}

\usepackage{graphicx}

\usepackage{relsize} 

\DeclareFontFamily{U}{mysize}{} 
\DeclareFontShape{U}{mysize}{m}{n}{ <-> s*[0.5] cmr10 }{} 

\DeclareMathOperator{\sinc}{sinc}

\usepackage{adjustbox}

\usepackage{xcolor}

\usepackage{caption}

\usepackage{subcaption}

\usepackage{comment}

\begin{document}

\preprint{APS/123-QED}

\title{Bound States in Continuum via Singular Transfer Matrices}

\author{Ovidiu-Zeno Lipan}%
 \email{olipan@richmond.edu}
\affiliation{
Department of Physics, Gottwald Center for the Sciences 138 UR Drive, University of Richmond, Richmond, VA 23173, USA
}%
\author{Aldo De Sabata}
\affiliation{
 Department of Measurements and Optical Electronics, Politehnica University of Timisoara, 300006 Timisoara, Romania}

\date{July 10, 2024}

\begin{abstract}
 
 In recent years, bound states in continuum (BICs) have gained significant value for practitioners in both theoretical and applied photonics. This paper focuses on devices that utilize non-homogeneous thin patterned laminae. The properties, design principles, and behavior of BICs for this class of devices are frequently explained through a variety of models, ranging from numerical or semi-analytical solutions for the Maxwell equations to heuristic approaches that rely on fitting functions to provide phenomenological descriptions. The field of devices under study has given less attention to approaches that integrate exact analytical solutions of the transfer matrix with numerical data.
 In this vein, this paper aims to adopt an approach where exact analytical formulas, detailed in our previous manuscript arXiv:2303.06765 (2023), are translated into equations to explore the origins and properties of the bound states in continuum, as well as their practical implementation. The geometric parameters of the device and its operating frequency band emerge from the null space of the transfer matrix.
\end{abstract}


\maketitle

\section{\label{sec:Introduction}Introduction}

The recent traction gained by the bound states in continuum (BICs) is exemplified by the hundreds of citations in \cite{hsu2016bound} and \cite{koshelev2023bound}, with \cite{koshelev2023bound} specifically containing a section dedicated to the history of the subject. In Section \ref{sec:ConclusionsDiscussions}, we  discuss some of these articles, establishing connections between our findings and this body of literature.

For a brief overview of where our process fits within the extensive landscape of BICs, we note that our approach begins with a set of experimentally relevant geometric and material devices, each distinguished by its unique parameters, as illustrated in Fig.\ref{fig:Device}.

The BIC states are obtained by analytically computing the transfer matrices of these varied devices,  as described in \cite{lipan2023closed}. Subsequently, each matrix is driven into a singular point, as shown in (\ref{eq:Sinc}), (\ref{eq:ConstrainOmegaForThetaZero}), Fig.\ref{fig:SurfaceThetaZeroViewPoint2} and
Fig.\ref{fig:EtaXiContourPlotThetaZeroLegended}.
At this singular point, the transfer matrix is not invertible, and thus the scattering matrix lacks clear definition. 
This singularity is linked to BIC states, which correspond to the null space of the transfer matrix at the singular point,  as emphasized in the discussions of formulas (\ref{MatrixBIC}) and (\ref{MatrixResonance}). 
We identified two frequency regions: one that hosts a pair of BICs, and another where a resonance and a BIC coexist.
In the vicinity of such a singular point, the scattering matrix is well defined (\ref{Sw1w1}) to  (\ref{Sw4w1}), giving rise to quasi-BIC states (\ref{eq:quasi BIC}). We explore the trajectories in the complex frequency plane of zeros and poles of the scattering matrix elements as the quasi-BIC regime transitions into a BIC, Fig.\ref{fig:NonCommutingLimitsWithPaths}. Finally, we study the robustness of the findings as the dimension of the linear space of modes, and consequently the dimension of the transfer matrix, increases, Fig.\ref{fig:DetCoff11ManyModes}.
The predicted properties of the BICs, some of which were computed using  \cite{Mathematica}, are validated through full-wave numerical simulations \cite{studios2022cst}, as seen in Fig.\ref{fig:All3BICSandResonance}.
The framework outlined in this manuscript, which involves creating BICs using a singular transfer matrix, is based on a newly developed method \cite{lipan2023closed}, indicating the potential for this novel approach to offer valuable insights and practical applications.

It is customary to start a photonics paper on bound states in continuum by citing the significant paper  by von Neumann and Wigner on unusual discrete eigenvalues \cite{vonNeumann:1929:MDE}. The process they use to  arrive at the potential energy leading to the desired uncommon eigenvalue follows a specific constructive methodology outlined in reference \cite{stillinger1975bound}. The process begins by solving the Schrödinger equation for the potential energy, assuming a known eigenstate. To apply a similar process to light propagation, we draw an analogy between the Hamiltonian, which governs temporal state transitions, and the transfer matrix, which governs electromagnetic state transitions but in space \cite{tomomaga1966quantum}, \cite{koshelev2023bound}. 
In light of the observed similarities, our objective is to devise a constructive method that determines the geometrical parameters and permittivities by imposing specific constraints on the transfer matrix to arrive at bound states in continuum (\ref{eq:T23T22}), (\ref{eq:T23T45}).

Whereas the quantum definition of the bound state in the continuum is clear from the beginning in  \cite{vonNeumann:1929:MDE}, applying the same concept to photonic devices in general proves more challenging \cite{hsu2016bound}. For example, take an extended periodic structure in $x$ and $y$ directions  but with a finite thickness in $z$ direction. Precise definition of BIC states is crucial within these systems, as is explicitly pointed out in \cite{hsu2016bound}. 
 
Instead of starting with a definition, the BIC states and their properties will gradually emerge in this paper as we explore the consequences derived from the intentionally constructed null space of the transfer matrix.

To materialize the transfer matrix and to illustrate the point above, we  had the option to utilize various polygonal shapes to construct a device for study, such as the one depicted in Fig.1 of \cite{Lipan:24}.  Nevertheless, we opted for the device shown in Fig.\ref{fig:Device}, given its frequent utilization \cite{wang2023brillouin}. The device is finite along the $z$-axis, having a thickness denoted by $g$, but periodically infinite in the $x$ and $y$-directions. 

The transfer matrix maps the electromagnetic field from location $z$-min to the field at location $z$-max. We placed $z$-min and $z$-max $g/4$ units away from the device in the  negative and positive  $z$-directions, respectively, Fig.\ref{fig:Device}.  The specific positions of $z$-min and $z$-max are connected to the space discretization proposed in \cite{pendry1992calculation}. Specifically, the space is viewed as a discrete lattice with lattice constants $(a,b,c)$ along $(x,y,z)$,  respectively. In \cite{lipan2023closed} we took the limit $a,b\rightarrow 0$, so only the $z$-direction remains discrete. In the context of a device, the lattice constant $c$  represents the thickness of one of the device's laminae. 

The device solved for in reference \cite{lipan2023closed} is comprised of two distinct laminae. However, in this paper, we assume the laminae to be identical, each with a thickness of $c$ resulting in an overall thickness of $g=2c$ for the device in Fig.\ref{fig:Device}.

 The transfer of the electromagnetic field from one plane located at $z=n c$ to $z=(n+1)c$, where $n$ is an integer, is done by two operators in two steps: first, the E-field is pushed forward from $n c$ to $nc+c/2$, followed by a second operator which transfers the H-field from $n c+c/2$ to $(n+1)c$. 
This is the reason why we placed 
$z_{\text{min}}$ at the position  $c/2=g/4$, in vacuum, away from one side of the device, and finished the transfer at $z_{\text{max}}$ placed at $c/2=g/4$, also in vacuum, but on the other side of the device.



The labeling structure of the transfer matrix is hierarchically nested, with each level contained within the one above \cite{lipan2023closed}. The first level reveals a 4x4 grid where each block exhibits  the matrix elements that couple between the left $(-)$ and right $(+)$ propagation directions along the z-axis

\begin{equation}\label{BlockMatrix}
T = \begin{pmatrix}
T^{+,+} & T^{+,-} \\
T^{-,+} & T^{-, -} 
\end{pmatrix}.
\end{equation}

The remaining labels associated with each element of the $T^{\pm,\pm}$-matrix specify the polarization and the Bloch-Floquet mode (BF) defined as in \cite{lipan2023closed}. Given the periodicity of the device in both the $x$ and $y$ directions, the transmitted and reflected fields can be expressed as a superposition of BF modes $(M_x,M_y)$, where $M_x$ and $M_y$ can take any integer value.
Throughout the paper, we will confine our discussion to TE polarization since the TM modes do not couple significantly into the TE.  Consequently, the elements of each block matrix in (\ref{BlockMatrix})  are identified by their respective BF indices.

The transmission scattering matrix, which takes input from location $z$-max and produces output at location $z$-min, requires the inversion of the $T^{-,-}$ matrix

\begin{equation}\label{eq:Sminmax}
    S^{min,max}=(T^{-,-})^{-1}.
\end{equation}

Therefore, studying the singularities in transmission provides a natural starting point.

The computational process might appear deceptively simple at first glance. However, the ability of this approach to distill intricate analytical formulas into a concise form may be unexpected if not surprising.



\section{\label{sec:THETA_Zero} Generating a singular transfer matrix}

\begin{figure}
    \includegraphics[width=0.75\linewidth]{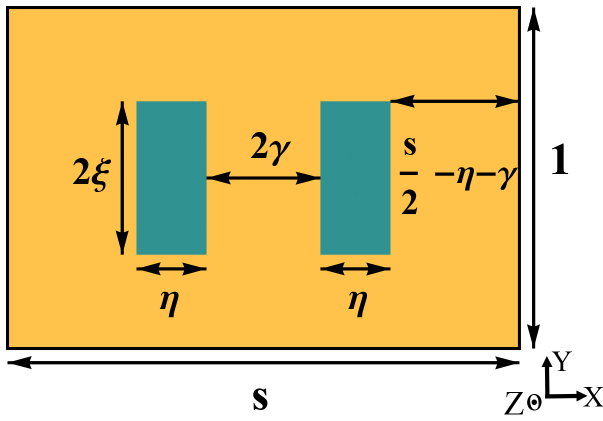}
\captionsetup{justification=raggedright,singlelinecheck=false}
    \caption{To illustrate the core concepts, we have chosen this type of geometry, although any bilaminar tessellated polygonal shape could be used. 
    All the geometrical data are unitless, scaled by $l_y$, the unit cell's vertical edge. To obtain the geometry in the chosen units, multiply all geometrical parameters by $l_y$.
    The two holes, symmetrically placed with respect to the center of the unit cell, have relative permittivity $\epsilon=1$, whereas the surrounding is silicon of $\epsilon_B=11.9$.  The unit cell is not necessarily a square, which is evident when $s\neq 1$. In this study, we opt to work with $s=1$.  We will then explore the case $s\neq1$ in a separate publication. } 
    \label{fig:Device}
\end{figure}

The incident plane wave, characterized by the BF mode 
$(0,0)$, travels from $z$-max to $z$-min. Its propagation direction is defined by the polar angle $\theta$ and the azimuthal angle  $\varphi$, oscillating at an angular frequency $\omega$.
In all the formulas that follow, the angular frequency is represented by a unitless parameter $\Omega$ defined as  $\omega/v_0=\Omega/c$, so $\Omega=(g/2)(\omega/v_0)$, where $v_0$ is the speed of light in vacuum.

In the approach described below, we first ensure that the projection of the transfer matrix  $T^{-,-}$ onto a lower-dimensional subspace spanned by a particular selection of BF modes is singular.
Subsequently, in Sec.\ref{sec:higher modes}, we assess whether the BICs formed within a lower-dimensional subspace persist within a higher-dimensional one.  

The lower-dimensional subspace employed is five-dimensional, with the BF modes ordered and numbered from 1 to 5 as follows: $(0,0),(-1,0),(1,0),(0,-1)$, and $(0,1)$. There is freedom in selecting the dimension of the lower-dimensional subspace, depending on the number of BICs we need to obtain. It turns out that we generate three BICs within this five-dimensional space.

There is a particular circumstance to consider when the angle $\theta$ 
approaches zero due to the specific structure of the BF basis we use. In particular, the vector 
$(E_x,E_y,H_x,H_y)$, which represents the state of the electromagnetic field proposed in \cite{pendry1992calculation} and adopted in \cite{lipan2023closed}, has 
$E_z$ and $H_z$ left aside, as they are dependent on the other four components.
As a consequence, the first vector of the BF basis, which is associated with the incoming mode $(0,0)$, tends toward zero as $\theta$ diminishes. This happens because all of its components share a common factor of $\sin(\theta)$. To avoid working with a zero vector, we perform a change of basis by employing the subsequent diagonal matrix $U=\text{diag}(\sin (\theta ),1,1,1,1)$
\begin{equation}\label{new basis}
    T^{-,-}|_{\text{new basis}}=UT^{-,-}U^{-1}.
\end{equation}
 In what follows we will not carry on the label "new basis" in the context of $\theta=0.$

By employing the method presented in \cite{lipan2023closed} to compute the analytical formulas for the transfer matrix we derive the following structure
\begin{equation}\label{TmatrixThetaZero}
T^{-,-}=\begin{pmatrix}
  T_{11} & T_{12} & -T_{12} & 0 & 0 \\
  T_{21} & T_{22} & T_{23} & 0 & 0 \\
  -T_{21} & T_{23} & T_{22} & 0 & 0 \\
  0 & 0 & 0 & T_{22} & T_{45} \\
  0 & 0 & 0 & T_{45} & T_{22} 
\end{pmatrix}   
\end{equation}
for the device from Fig.\ref{fig:Device} at $\theta=0,\varphi=0$ and $s=1$, making the unit cell of this thin device a square for the rest of this paper.
The specific format outlined by (\ref{TmatrixThetaZero}) represents the manifestation of symmetry in the transfer matrix pertinent to the device class depicted in Fig.\ref{fig:Device} with $s=1$. Symmetries  related to the $\Gamma$-point are ubiquitous. 

Driving the transfer matrix to a singular point necessitates additional constraints. Here, we propose to request that, at  $\theta=0$ and $\varphi=0$, there exists a frequency $\Omega$  located on the real axis of the complex frequency plane such that
\begin{align}\label{eq:DetCof}
\text{det}(T^{-,-})=0\\\label{eq:CofZero}
\text{Cof}(T^{-,-})_{11}=0
\end{align}
where $\text{Cof}(T^{-,-})_{11}$ is the cofactor of the element $T^{-,-}_{11}$ from the transfer matrix.

The reason behind this choice is that, for a well-defined  scattering matrix, (\ref{eq:Sminmax}), the transmission of the  incoming propagative mode $(0,0)$ is
\begin{equation}\label{eq:SMinMax}
    S^{min,max}_{11}=\frac{\text{Cof}(T^{-,-})_{11}}{\text{det}(T^{-,-})}.
\end{equation}

Through (\ref{eq:DetCof}) and (\ref{eq:CofZero}), we establish the possibility for the transmission to have a finite, non-zero value in the limit of $\theta \rightarrow 0$, akin to L'Hôpital's rule. Away from  $\theta=0$ where the cofactor and the determinant no longer share a common root, the transmission will be extinguished by the zero of the cofactor.
The imposed conditions suggest the presence of a BIC where the transmission abruptly transitions from being zero at $\theta\neq 0$ to a non-zero value close to 1 at $\theta=0$. This represents a   key point in introducing a relevant singularity for generating BICs.


These two constraints typically necessitate their separation into real and imaginary parts, as our matrix elements are generally complex numbers. However, by exploiting the fact that transfer matrix elements are real numbers for the evanescent modes in lossless dielectric devices, we can reduce the number of equations needed to impose the singularity.

Out of several alternatives to satisfy (\ref{eq:DetCof}) and (\ref{eq:CofZero}), we chose two relationships among the matrix elements
\begin{eqnarray}\label{eq:T23T22}
   T_{23}=T_{45}\\\label{eq:T23T45}   
      T_{23}=-T_{22}
\end{eqnarray}
that not only ensure the validity of (\ref{eq:DetCof})  and (\ref{eq:CofZero}) but also result in a shape for $T^{-,-}$
\begin{equation}\label{eq:H-H}
T^{-,-}=\begin{bmatrix}
  T_{11} & T_{12} & -T_{12} & 0 & 0 \\
  T_{21} & H & -H & 0 & 0 \\
  -T_{21} & -H & H & 0 & 0 \\
  0 & 0 & 0 & H & -H \\
  0 & 0 & 0 & -H & H 
\end{bmatrix}   
\end{equation}
that sustains a double real root in frequency for both the determinant and cofactor, which will produce two BICs. The constraint (\ref{eq:T23T45}) extends the already existing relation $T_{13}=-T_{12}$, whereas (\ref{eq:T23T22})  duplicates the 2x2 matrix of the BF modes 2 and 3 to the pair 4 and 5.


We have reached the decisive moment to leverage the analytic closed-form formulas for the transfer matrix elements, enabling us to distill the first equation (\ref{eq:T23T22}) into
\begin{equation}\label{eq:Sinc}
\xi \sin(4 \pi \gamma)-\xi \sin(4 \pi (\gamma+\eta))+\eta \sin(4 \pi \xi)=0
\end{equation}
 whereas (\ref{eq:T23T45}) gives
\begin{equation}\label{eq:ConstrainOmegaForThetaZero}
    \eta \xi (1-\sinc(4\pi\xi))=\frac{1}{4}\frac{((\epsilon_B-1)\Omega^2 \Phi_e-1)^2-(\Phi_e)^2}{((\epsilon_B-1)\Omega^2 \Phi_e-1)^2-1}
\end{equation}
with $\sinc(x) =x^{-1} \sin(x)$.
Here $\Phi_e$ captures the propagation along the $z$-axes of evanescent waves, as described in \cite{lipan2023closed}
\begin{equation}
    \Phi_e=\Phi_{Z,\text{evanescent}}\left(\frac{1}{2}(2+w^2-\Omega^2)\right)
\end{equation}
with 
\begin{eqnarray}\label{eq:PHIe}
     \Phi_{Z,\text{evanescent}}(z)&=&z-\sqrt{z^2-1}\\\label{eq:w}
     w&=&g \frac{ \pi}{l_y}.
\end{eqnarray}

The formulas above incorporate the $w$-parameter, which depends on both the unit cell's $y$-axis length ($l_y$) and its thickness along the z-axis, ($g$). In this manuscript, when designing a specific device, we start by setting a value for $w$ by choosing $l_y$ and $g$, with $g\ll l_y$ fulfilling the requirement for thin laminae.  Then, the three geometrical parameters  $(\eta,\xi,\gamma)$ are chosen to satisfy (\ref{eq:Sinc}) and, finally, the frequency $\Omega$ is obtained from (\ref{eq:ConstrainOmegaForThetaZero}).  However, the process of imposing a singularity can begin with specifying the frequency first, which is equally valid.

For  numerical computations that follow, we choose $l_y=120 \,\mu\text{m}$ and $g=2\,\mu\text{m}$.

\begin{figure}
    \centering
    \includegraphics[width=0.75\linewidth]{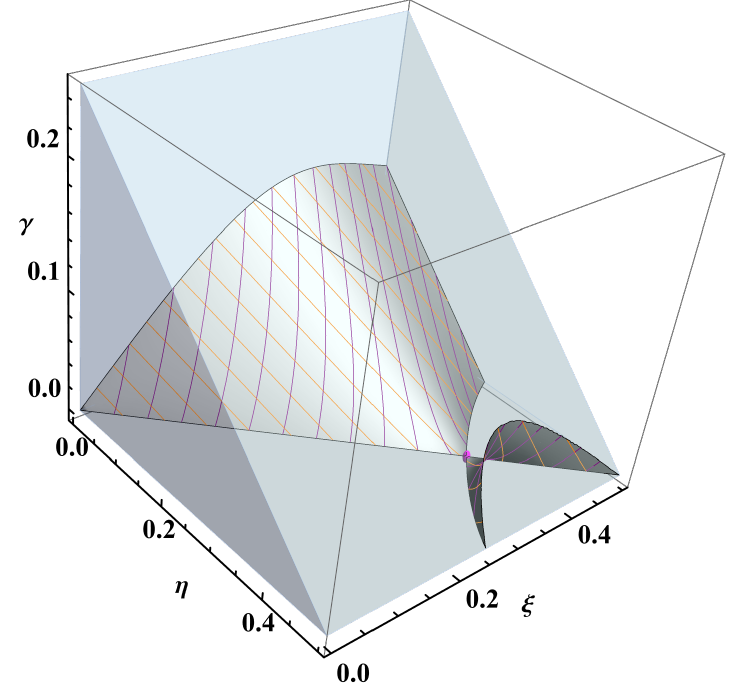}
\captionsetup{justification=raggedright,singlelinecheck=false}
    \caption{The 3D surface for the BICs devices that fulfill the shape (\ref{eq:H-H}) and have a square unit cell, $s=1$. Note that this surface does not depend on $w$ from (\ref{eq:w}).}
    \label{fig:SurfaceThetaZeroViewPoint2}
\end{figure}
Equation (\ref{eq:Sinc}) defines a constraint that represents the surface shown in Fig.\ref{fig:SurfaceThetaZeroViewPoint2}. Due to the device periodicity in the $x$-$y$ plane, we restrict the plotted region using inequality  $s/2-\eta-\gamma \geqslant\gamma$ to avoid including the same device multiple times.

This surface consists of two separate surfaces that touch at a point where  $\gamma=0$, signifying the merging of two distinct holes into one.
At this point,
\begin{equation}\label{eq:Eye}
    \eta=\xi=0.3575741632810507.
\end{equation} 
However, this isn't the sole point in the geometric space $(\eta,\xi,\gamma)$ where $\gamma=0$ holds true. Equation 
\begin{equation}\label{eq:SincGammaZero}
\sinc(4 \pi \xi)=\sinc(4 \pi \eta)
\end{equation}
is valid for all such points, characterized by the diagonal $\eta=\xi$ and an additional curve visible in the $\gamma=0$ plane in Fig.\ref{fig:SurfaceThetaZeroViewPoint2}. 
We will name the point (\ref{eq:Eye}) 'SincSync' in recognition of its position at the intersection of two distinct curves associated with the equation $\sinc(4 \pi \xi)=\sinc(4 \pi \eta)$. 

\begin{figure}
    \includegraphics[width=0.75\linewidth]{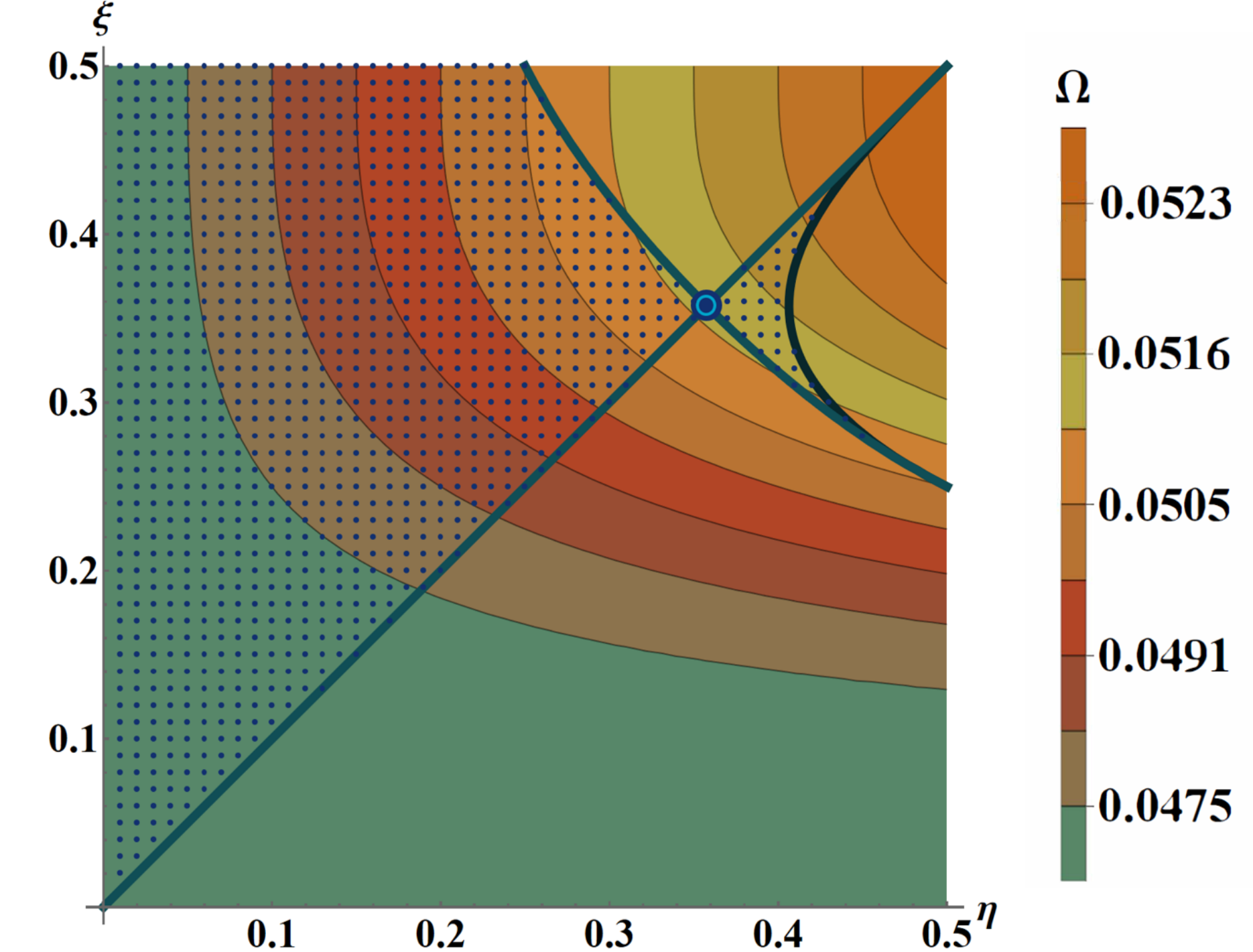}  \captionsetup{justification=raggedright,singlelinecheck=false}
    \caption{The dotted region is the projection of the surface from Fig.\ref{fig:SurfaceThetaZeroViewPoint2} on the plane $(\eta,\xi))$. The hyperbolic-like contour plots represent formula (\ref{eq:ConstrainOmegaForThetaZero}), i.e. they give the constant frequency lines. The legend might have been labeled based on the values of $\eta \xi (1-\sinc(4\pi\xi))$, from (\ref{eq:ConstrainOmegaForThetaZero}). However, it is instead assigned the frequency $\Omega$ obtained from the same formula for $l_y=120\,\mu\text{m}$ and $g=2\,\mu\text{m}$, facilitating a straightforward correlation with the plots from the numerical example. }
\label{fig:EtaXiContourPlotThetaZeroLegended}
\end{figure}

After selecting $(\eta,\xi,\gamma)$ from the surface depicted in Fig.\ref{fig:SurfaceThetaZeroViewPoint2}, the second constraint (\ref{eq:ConstrainOmegaForThetaZero}) determines the frequency required to achieve the desired pattern for the transfer matrix (\ref{eq:H-H}). 
The contour lines of constant frequency are shown in 
 Fig.\ref{fig:EtaXiContourPlotThetaZeroLegended}, together with the projection of the surface from Fig.\ref{fig:SurfaceThetaZeroViewPoint2} onto the plane $(\eta,\xi)$. Besides the two curves from (\ref{eq:SincGammaZero}), a third curve emerges from the downward projection.

In our  computations and numerical simulations, we selected the SincSync point, (\ref{eq:Eye}), clearly identified in Fig.\ref{fig:EtaXiContourPlotThetaZeroLegended}, for which the frequency is $\Omega=0.05126$, given that  $l_y=120 \,\mu\text{m}$ and $g=2\,\mu\text{m}$. Consequently, the transfer matrix (\ref{eq:H-H}) takes the numerical form of
\begin{equation}\label{eq:Numerical_At-T_antisymmetric}
\scalebox{0.70}{%
\ensuremath{
\begin{bmatrix}
0.947 - 0.419i & -0.008 - 0.101i & 0.008 + 0.101i & 0 & 0 \\
-0.466 + 0.036i & -0.304 & 0.304 & 0 & 0 \\
0.466 - 0.036i & 0.304 & -0.304 & 0 & 0 \\
0 & 0 & 0 & -0.304 & 0.304 \\
0 & 0 & 0 & 0.304 & -0.304 
\end{bmatrix}.
}
}
\end{equation}
\begin{figure}
    \includegraphics[width=1\linewidth]{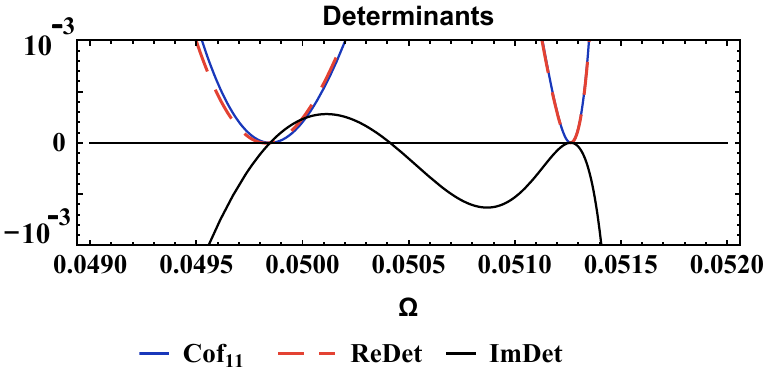}
\captionsetup{justification=raggedright,singlelinecheck=false}
    \caption{Determinants. ReDet and ImDet are the real and the imaginary  part of the determinant of the $T^{-,-}$-matrix based on the first five BF modes,  $(0,0),(-1,0),(1,0),(0,-1)$, and $(0,1)$, respectivelly. $\text{Cof}_{11}$ is the cofactor for the element $T^{-,-}_{11}$ of this matrix. A common zero of (\ref{eq:DetCof}) and (\ref{eq:CofZero}) is present at the designed singularity $\Omega_a=0.051263488625532705$. As a bonus, a second singularity appears at a lower frequency $\Omega_s=0.04984752722741948$. The ReDet and $\text{Cof}_{11}$ both have a parabolic shape, taking values above zero.}
    \label{fig:Det5by5}
\end{figure}

The behavior of the transfer matrix of the device away from $\Omega=0.05126$ is illustrated in Fig.\ref{fig:Det5by5}. A surprising observation is that at lower frequencies, the device exhibits a second frequency worthy of further study.

At this lower frequency, $\Omega=0.049847$, the transfer matrix is given by (\ref{eq:Numerical_At-Tsymmetric}), where the structure of the 4x4 submatrix derived from it\textemdash after  removing the first column and row\textemdash exhibits a symmetric pattern. This contrasts with the antisymmetric patterns shown in (\ref{eq:H-H}) and (\ref{eq:Numerical_At-T_antisymmetric}),
\begin{equation}\label{eq:Numerical_At-Tsymmetric}
\scalebox{0.70}{%
\ensuremath{
\begin{bmatrix}
0.950 - 0.408i & -0.008 - 0.102i & 0.008 + 0.102i & 0 & 0 \\
-0.288 + 0.022i & 0.194 & 0.194 & 0 & 0 \\
0.288 - 0.022i & 0.194 & 0.194 & 0 & 0 \\
0 & 0 & 0 & 0.194 & 0.194 \\
0 & 0 & 0 & 0.194 & 0.194 \\
\end{bmatrix}.
}
}
\end{equation}

The patterns are classified as antisymmetric and symmetric based on the action of the 4x4 submatrices on the BF modes, which yield antisymmetric and symmetric linear combinations of the BF  modes, respectively. The 4x4 submatrices can also be viewed as composed of 2x2 matrices that follow the same pattern as the Jones matrices for  $-45^{\circ}$ and  $+45^{\circ}$  ideal polarizers, respectively.

This numerical finding indicates the possibility that both patterns\textemdash the symmetric 4x4 submatrix structure
\begin{equation}\label{eq:HH}
\begin{bmatrix}
   H & H & 0 & 0 \\
   H & H & 0 & 0 \\
   0 & 0 & H & H \\
  0 & 0 & H & H 
\end{bmatrix}   
\end{equation}
and the antisymmetric one from (\ref{eq:H-H})\textemdash may coexist within the same device. These patterns  appear at different frequency values when scanning through the points across the surface depicted in Fig.\ref{fig:SurfaceThetaZeroViewPoint2}, where the points correspond to different devices.

\begin{figure}
    \centering
    \includegraphics[width=0.75\linewidth]{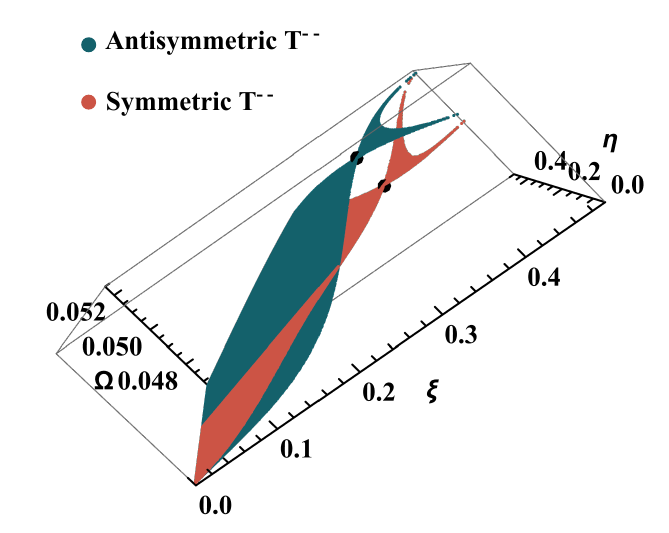} \captionsetup{justification=raggedright,singlelinecheck=false}
    \caption{Separate frequency surfaces correspond to the unitless frequency parameters $\Omega_a$ and $\Omega_s$, representing antisymmetric and symmetric patterns, respectively.
    The intersection of these two surfaces is along the segment $\xi=1/4$  and $0<\eta<1/4$. On this segment $\gamma=(1/8)-(\eta/2)$ and $H=0$.}
    \label{fig:Goanga}
\end{figure}

\begin{figure}
    \centering
    \includegraphics[width=0.75\linewidth]{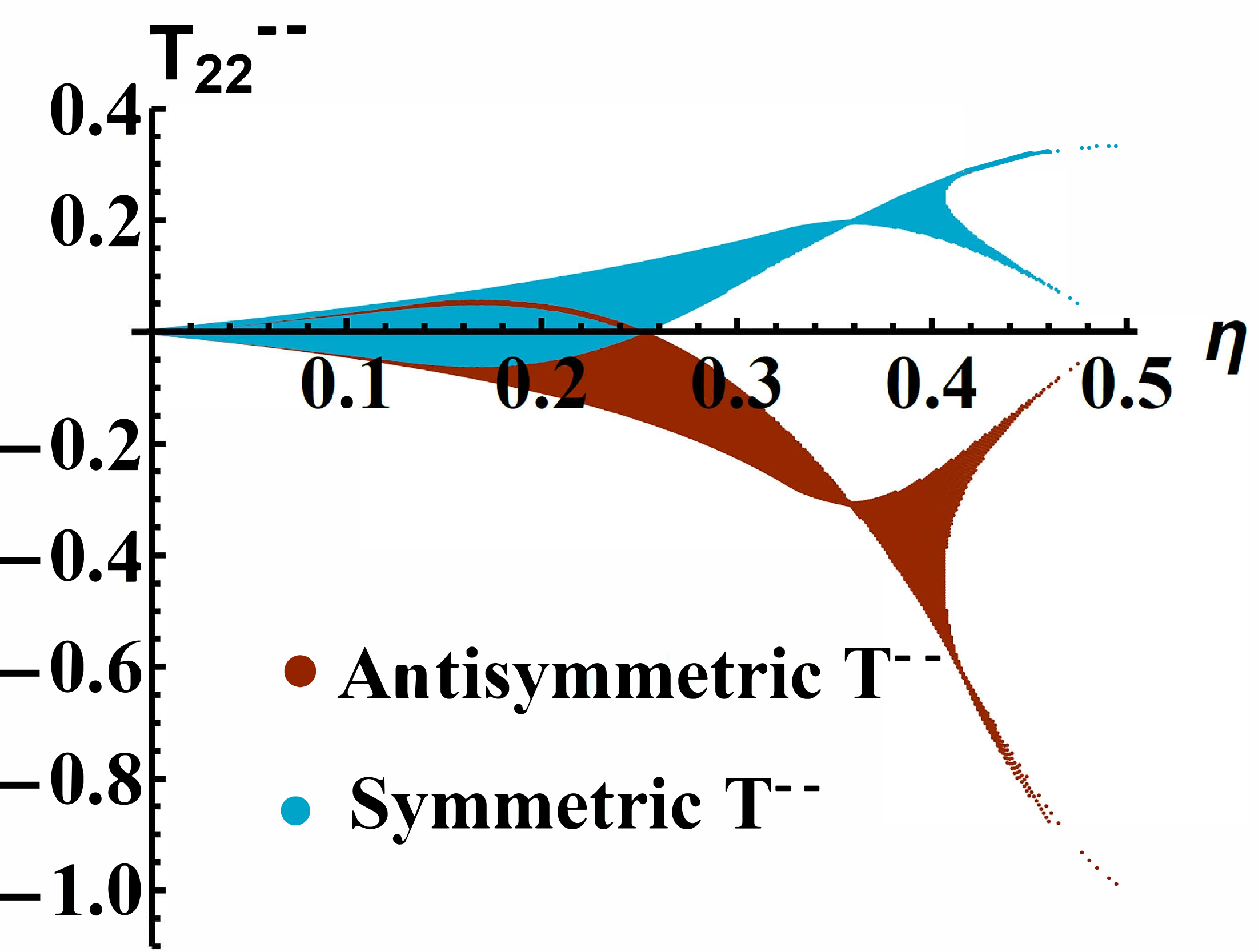}
\captionsetup{justification=raggedright,singlelinecheck=false}
    \caption{The values of $T^{- -}_{22}$ for the antisymmetric and symmetric  patterns. Above each value for $\eta$ the segment represents all values of  $T^{- -}_{22}$ for all possible values of $\xi$.}
    \label{fig:T22ValuesPeackokColorV2}
\end{figure}

Indeed, in our quest for a symmetric  $T^{- -}$ structure (\ref{eq:HH}), we find that while the first condition (\ref{eq:Sinc}) remains unchanged, the second condition transforms into 
\begin{equation}\label{eq:ConstrainOmegaForThetaZeroSymmetric}
    \eta \xi (1+\sinc(4\pi\xi))=\frac{1}{4}\frac{((\epsilon_B-1)\Omega^2 \Phi_e-1)^2-(\Phi_e)^2}{((\epsilon_B-1)\Omega^2 \Phi_e-1)^2-1}.
\end{equation}
Both symmetric and antisymmetric patterns require our attention going forward. As the patterns manifest at different frequencies, a natural inquiry arises: how far apart are these two frequencies as we explore all geometrical parameters?
Fig.\ref{fig:Goanga} captures this frequency difference. A notable distinction exists for the SincSync point compared to many others.
The comprehensive view of device behavior across various geometries, as illustrated in Fig.\ref{fig:Goanga}, can be expanded to encompass other parameters positioned along the vertical axis. Building on this idea, Fig.\ref{fig:T22ValuesPeackokColorV2} clarifies the distinction in values taken by matrix element $H=T^{- -}_{22}$ for the symmetric and antisymmetric patterns.
Having selected the SincSync device and identified the frequencies of interest, our analysis diverges into two main areas. 

First, we establish the basis of the linear space where the relevant modes reside, providing the fundamental set of vectors necessary to discuss the resonances and BIC phenomena. Second, we move to the complex frequency plane to explore the trajectories of the zeros and poles of  the scattering matrix elements as $\theta$ and $\varphi$ vary.
The methodology we are imparting here holds versatility\textemdash it can be wielded across various points on the surface of Fig.\ref{fig:SurfaceThetaZeroViewPoint2}.

We base our five-dimensional linear basis on the four eigenvectors of the  4x4 submatrix of the antisymmetric and symmetric patterns, plus the vector $W_1$ that corresponds to the incoming $(0,0)$ BF mode
\begin{align}
W_1=(1,0,0,0,0),\; W_2=(0,1,1,0,0)/\sqrt{2}\\\nonumber
W_3=(0,-1,1,0,0)/\sqrt{2},\;W_4=(0,0,0,1,1)/\sqrt{2}\\\nonumber
W_5=(0,0,0,-1,1)/\sqrt{2}.
\end{align}


In this W-basis, the $T^{-,-}$-matrix (\ref{eq:H-H}), which exhibits the antisymmetric pattern, becomes
\begin{equation}\label{MatrixBIC}
\begin{bmatrix}
 T_{11} & 0 & -\sqrt{2} T_{12} & 0 & 0 \\
 0 & 0 & 0 & 0 & 0 \\
 -\sqrt{2} T_{21} & 0 & 2 H & 0 & 0 \\
 0 & 0 & 0 & 0 & 0 \\
 0 & 0 & 0 & 0 & 2 H \\
\end{bmatrix}
\end{equation}
whereas for the symmetric pattern  (\ref{eq:HH}), the $T^{-,-}$-matrix becomes
\begin{equation}\label{MatrixResonance}
\begin{bmatrix}
 T_{11} & 0 & -\sqrt{2} T_{12} & 0 & 0 \\
 0 & 2 H & 0 & 0 & 0 \\
 -\sqrt{2} T_{21} & 0 & 0 & 0 & 0 \\
 0 & 0 & 0 & 2 H & 0 \\
 0 & 0 & 0 & 0 & 0 \\
\end{bmatrix}.
\end{equation}

At the antisymmetric pattern frequency $\Omega_a$, Fig.\ref{fig:Det5by5}, the matrix structure  (\ref{MatrixBIC}) predicts that the device has two BICs corresponding to $W_2$ and $W_4$, as observed in the all-zero columns 2 and 4. The association of BICs with zero eigenvalues is also noted in \cite{shubin2023algebraic}.

At the symmetric pattern frequency, $\Omega_s$,  where the structure is given by (\ref{MatrixResonance}), $W_3$ is a resonance and $W_5$  a BIC, evident from columns 3 and 5.
To further confirm that $W_3$ is a resonance at the lower frequency $\Omega_s$, we examine the 2 by 2 matrix coupling $W_1$ and $W_3$ in (\ref{MatrixResonance}). Its elements exhibit the resonant trait: in this particular W-basis, $T_{33}$ equals zero, while the remaining three elements do not \cite{png2017nanophotonics}. 

The distinction between the resonance and the BIC, along with further clarification of their meanings, is fundamentally related to the zeros and poles of the scattering matrix elements, as will be covered in Section \ref{sec:ResonanceBIC}.




In the forthcoming sections, we delve into these predictions. Of particular interest is the conversion of a BIC into a quasi-BIC, necessitating the transition from $\theta=0$ to $\theta \neq 0$.


\section{Analysis of the two BICs at the antisymmetric pattern frequency $\Omega_a$. }


To investigate the behavior of the device away from $\theta=0,\varphi=0$ and within frequency regions centered on  $\Omega_{a/s}$, we employ a series expansion of the transfer matrix elements expressed in the BF basis. It is not obvious upfront if a first order in $(\theta,\varphi)$ works, so we went with a second order expansion to hopefully capture the behavior of the BICs
\begin{align}\label{eq:Tseries}
T^{-,-}=& T_{000} + T_{010}\; \theta + T_{001}\; \varphi + T_{020}\; \theta^2 \\\nonumber
& + T_{011}\; \theta \varphi + T_{002}\; \varphi^2 + T_{100}\; \Delta +\cdots
\end{align}
Here $\Delta$ is a small variation in frequency away either from the antisymmetric $\Omega_{a}=0.0512$ or symmetric frequency $\Omega_{s}=0.0498$
\begin{eqnarray}
    \Omega=\Omega_{s/a}+\Delta_{s/a}.
\end{eqnarray}
This expansion was carried out  in both symbolic and numerical format. As a result, the determinant of $T^{-,-}$ and all its cofactors become polynomials of at least fourth order in the frequency parameter $\Delta_{s/a}$. 
Given that the observed behavior resembles a parabolic shape in Fig.\ref{fig:Det5by5}, we require only second-order rational functions in $\Delta_{s/a}$ for each scattering matrix element. 
In other words, two zeros and two poles for each of the scattering matrix elements will be both necessary and sufficient to accurately represent the behavior of either the pair of BICs at  $\Omega_a$ or the pair consisting of a resonance and a BIC at $\Omega_s$. This conclusion explains why stopping at the first order in $\Delta$  in (\ref{eq:Tseries}) is sufficient.

While a series expansion around $\Delta_{s/a}=0$ is straightforward and useful, it might not provide a revealing understanding. We seek analytic formulas that not only yield results but also shed light on the underlying mechanisms.

The key to this approach is to identify those specific frequencies where the diagonal elements of the transfer matrix become zero. Fortunately, the formulas for these frequencies are readily available. Our goal then becomes building the poles and zeros of  the scattering matrix elements as perturbations of these frequencies.

This plan offers two advantages. First, we can treat the frequency region containing the two BICs independently from the resonance and its accompanying BIC. Second, this separation allows us to work with smaller, 3x3 submatrices.
One submatrix deals with the coupling of vectors $W_1, W_2$, and $W_4$ in relation to the two BICs, while the other concentrates on the coupling of $W_1, W_3$, and $W_5$ associated with  the resonance and its BIC.


We start by selecting the 3 by 3 submatrix that couples $W_1,W_2$ and $W_4$, i.e. the coupling between the incoming wave and the two BICs, where $\Delta_a=\Omega-\Omega_{a}$. The result is the matrix $T^{-,-}_{W_1W_2W_4}$ from (\ref{MatrixTw1w2w4}) expressed in the W-basis
\begin{widetext}

  \begin{minipage}{1\linewidth} 
    \begin{equation}\label{MatrixTw1w2w4}
   T^{-,-}_{W_1W_2W_4}= \begin{pmatrix}
  {A_{1 1} + \Delta_a \beta_{1 1} + \theta^2 \mu_{1 1}} & {\sqrt{2} \theta B_{1 2}} & {\sqrt{2} \theta B_{1 4}} \\
  {\sqrt{2} \theta B_{2 1}} & {\Delta_a \beta_{2 2} + \Delta_a \beta_{3 2} + \theta^2 \mu_{2 2} + \theta^2 \mu_{3 2}} & {2 \theta^2 \mu_{2 4}} \\
  {\sqrt{2} \theta B_{4 1}} & {2 \theta^2 \mu_{4 2}} & {\Delta_a \beta_{2 2} + \Delta_a \beta_{3 2} + \theta^2 \mu_{4 4} + \theta^2 \mu_{5 4}}
\end{pmatrix}.
\end{equation}
  \end{minipage}


\end{widetext}

The meanings of the symbols can be deduced from the accompanying parameters. For instance, the Greek symbol $\mu$ represents the matrix elements of $T_{020}$, reflecting the second-order expansion in $\theta$ as shown in (\ref{eq:Tseries}).
The scattering matrix elements are obtained by taking the inverse of 
$ T^{-,-}_{W_1W_2W_4}$, with the index 
$(min,max)$ implied,  whereas  in (\ref{eq:Sminmax}) it is explicit.

The poles of the scattering matrix elements are complex numbers, a property inherited from the elements in the first row and first column, which are complex numbers as they exhibit the coupling into the propagative BF mode $(0,0)$. 

In contrast, the zeros of $S_{W_1W_1}$, $\text{zero}_1$ from (\ref{eq:Zero1For2BICs}), and $\text{zero}_2$ from (\ref{eq:Zero2For2BICs}) are located on the real axis. This happens as they are based on transfer matrix elements that couple evanescent modes within themselves, which are real numbers in lossless dielectric devices.
Outside the bounds of the present approximation, when the number of modes is increased or the angles $\theta$ and $\varphi$
are no longer very small, the zeros start to move into the complex plane away from the real axis.

Before covering the steps we took to locate the zeros and poles, we present the results in (\ref{Sw1w1}), (\ref{Sw2w1}) and (\ref{Sw4w1}) to help assess the structure of the relevant scattering matrix elements in W-basis
\begin{align}\label{Sw1w1}
  S_{W_1W_1}&= \frac{1}{A_{11}}\,\underbrace{\frac{\Delta_a-\text{zero}_1}{\Delta_a-\text{pole}_{1}}\;}_{\text{quasi-BIC}_1}\;\underbrace{\frac{\Delta_a-\text{zero}_2}{\Delta_a-\text{pole}_{2}}}_{\text{quasi-BIC}_2}\\\label{Sw2w1}
   S_{W_2W_1}&=\theta \frac{-\sqrt{2} \,  B_{2 1}  }{A_{11} \left( \beta_{22} + \beta_{32} \right)} \;\frac{\Delta_a-\text{zero}_{W_2}}{(\Delta_a-\text{pole}_{1})(\Delta_a-\text{pole}_{2})}\\\label{Sw4w1}
   S_{W_4W_1}&=\theta \frac{-\sqrt{2} \,  B_{4 1} }{A_{11} \left( \beta_{22} + \beta_{32} \right)} \;\frac{\Delta_a-\text{zero}_{W_4}}{(\Delta_a-\text{pole}_{1})(\Delta_a-\text{pole}_{2})}.
\end{align}

The format of these matrix elements is that of a second-order rational function, which occurs in the field of Circuits and Systems \cite{roberts1991general}. Additionally they are utilized for fitting experimental data  in the field of Optics \cite{gibson2021pole},\cite{neviere1995electromagnetic}.

There are certain aspects of this result that are important to discuss. One aspect pertains to the structure of 
$S_{W_1W_1}$ when viewed as a product of two linear fractional functions, also known as M\"obius  or bilinear transformations, each representing a Fano line shape \cite{avrutsky2013linear}. This contrasts with the more common approach of considering the numerator and denominator separately, as second order polynomials \cite{roberts1991general}. Viewing 
$S_{W_1W_1}$ as a product of Fano line shapes brings us closer to the Weierstrass factorization for an entire function, which is employed in scattering problems \cite{grigoriev2013optimization},\cite{newton2013scattering},\cite{mikheeva2023asymmetric}. The study of BICs is facilitated by examining each linear fractional function, where its zeros and poles converge, as $\theta\rightarrow 0$, to a common point on the real axis.
The implications of these results will be further explored in Section \ref{sec:QuasiBIC}.
 Until then, we will go over the derivation of these scattering matrix elements.

First thing to be noticed, to isolate the zeros and poles, is that the diagonal matrix elements from position $(2,2)$ and $(3,3)$ present two distinct roots for $\Delta_a$ which are closed to each other (since $\theta$ is small)
\begin{align}\label{roots24}
    \text{root}_{W_2}=- \theta^2\,\frac{ \mu_{22} + \mu_{32}}{\beta_{22} + \beta_{32}}=-\frac{29}{18}\theta^2\\\label{roots44}
    \text{root}_{W_4}=- \theta^2\,\frac{ \mu_{44} + \mu_{54}}{\beta_{22} + \beta_{32}}=\frac{1}{44}\theta^2.
\end{align}
We use the term 'root' for the value of the first approximation and 'zero' or 'pole' for the  form with corrections included.
We included the numerical values in rational number format to make it visually easier to see the source of the various corrections that follow. This format is also useful for visualizing which numerical factors are close to each other and which factors simplify.
After expanding the determinant and the cofactors of $T^{-,-}_{W_1W_2W_4}$ in the vicinity of (\ref{roots24}) and (\ref{roots44}), the four zeros and the two poles of the scattering matrix elements will emerge as corrections to these roots:
\begin{align}
 \text{zero}_{W2}= \text{root}_{W4}+\theta^2\, \frac{2  B_{4 1} \mu_{2 4}}{B_{2 1} \left( \beta_{2 2} + \beta_{3 2} \right)}\\
 \text{zero}_{W4}= \text{root}_{W2}+\theta^2\, \frac{2  B_{2 1} \mu_{4 2}}{B_{4 1} \left( \beta_{2 2} + \beta_{3 2} \right)}.
\end{align}
The corrections are small, as illustrated by the numerical values shown below
\begin{align}\label{eq:ZeroW2}
\text{zero}_{W2}=\frac{1}{44} \theta^2-\left(\frac{1}{1563}+\frac{i}{11526773}\right)\theta^2\\\label{eq:ZeroW4}
 \text{zero}_{W4}=-\frac{29}{18}\theta^2 +\left(\frac{1}{45}-\frac{i}{335723}\right)\theta^2.
\end{align}

In contrast to  $\text{zero}_1$   and  $\text{zero}_2$ of $S_{W_1W_1}$, which are real numbers in this approximation, the zeros of $S_{W_2W_1}$ and $S_{W_4W_1}$ have an imaginary component introduced by the coupling with the propagative mode through $B_{2 1}$ and $B_{4 1}$. In fact, a non-zero imaginary part for zeros is more commonly encountered than its absence. 
The goal is to suppress the imaginary part as much as possible, which is achieved in (\ref{eq:ZeroW2}) and (\ref{eq:ZeroW4}).

The symbolic formulas for the poles are presented in  Appendix \ref{AppendixPolesAntisymmetric}. However, the numerical values are provided for comparison to the zeros and to highlight their nature as complex numbers.
\begin{align}
  \text{pole}_{1}=  -\frac{29}{18}\theta^2+\left(\frac{1}{650} - \frac{i}{173}\right) \theta^2\\
  \text{pole}_{2}=\frac{1}{44}\theta^2+\left(\frac{1}{27891} - \frac{i}{6126}\right) \theta^2 
\end{align}
\begin{widetext}

  \begin{minipage}{0.9\linewidth} 
\begin{align}\label{eq:Zero1For2BICs}
  \text{zero}_1= - \frac{ \mu_{2,2} +  \mu_{3,2}}{\beta_{2,2} + \beta_{3,2}}\theta^2 - \frac{4  \mu_{2,4} \mu_{4,2}}{(\beta_{2,2} + \beta_{3,2})(\mu_{2,2} + \mu_{3,2} - \mu_{4,4} - \mu_{5,4})}\theta^2=-\frac{29}{18} \theta^2 - \left(-\frac{\theta^2}{116197}\right)\\\label{eq:Zero2For2BICs}
  \text{zero}_2=-\frac{\mu_{4 4} + \mu_{5 4}}{\beta_{2 2} + \beta_{3 2}}\theta^2  + \frac{4  \mu_{2 4} \mu_{4 2}}{(\beta_{2 2} + \beta_{3 2})(\mu_{2 2} + \mu_{3 2} - \mu_{4 4} - \mu_{5 4})}\theta^2=\frac{1}{44} \theta^2 + \left(-\frac{\theta^2}{116197}\right).
\end{align}
\end{minipage}
\end{widetext}


Because $W_1=BF_1=(1,0,0,0,0)$, we have $S_{W_1W_1}=S_{11}|_{\text{base-BF}}$, which is directly visible in the numerical simulations, as shown in panels (a) to (c) of Fig.\ref{fig:All3BICSandResonance}. Additionally, note that we are working not only in the W-basis but also in a changed basis performed by the diagonal matrix $U$, (\ref{new basis}).
Keeping the W-basis and removing the $U$-dependent basis, we need to multiply $S_{W_2W_1}$ and $S_{W_4W_1}$ by $\sin(\theta)$ while $S_{W_1W_1}$ remains unchanged. This is important for comparing with the numerical simulations.
To avoid any confusion, we emphasize that we will continue to use the new basis (\ref{new basis}) in what follows.

\begin{figure}
    \centering
    \includegraphics[width=0.90\linewidth]{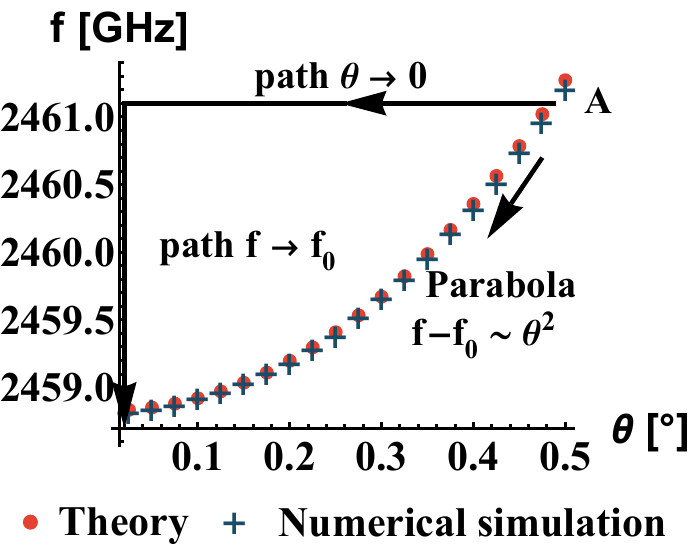}
\captionsetup{justification=raggedright,singlelinecheck=false}
    \caption{Following two distinct paths from point A toward the origin, we obtain different results, thereby justifying the non-commuting limits in (\ref{noncommuting_limits_S11}). The values of the pair $(\theta,f)$ that result in zero transmission $S_{W_1W_1}$ form a parabola. The set of frequencies corresponds to one of the two quasi-BICs located in the vicinity of $\Omega_a$. The quasi-BIC turns into a BIC for $\theta=0$ at $f_0=2458.8$ GHz. The translation between the frequency $f$ and the unitles frequency parameter $\Omega$ is $\Omega=(g/2)(2 \pi f/v_0)$, where $v_0$ is the speed of light in vacuum and $g$ the thickness of the device from Fig.\ref{fig:Device}. This parabola, as predicted by the theory for the chosen numerical values $l_y=120\mu\text{m}$ and $g=2\mu\text{m}$, is $f=2458.8 + 9.5\;\theta^2$ in GHz. As also reported in \cite{Lipan:24}, a systematic shift exists between the theoretical and simulated frequencies. Here, all the theoretical frequencies are 12.87 GHz less than the simulated ones. This amounts to a $0.52\%$ error for this frequency band. To enhance precision, we used a 9x9 transfer matrix to compute the parabola where the device is entirely opaque. The added four BF modes are $(\pm1,\pm1)$. The influence of higher modes is discussed in Sec.\ref{sec:higher modes}.  }
    \label{fig:NonCommutingLimitsWithPaths}
\end{figure}


\section{Quasi-BICs and non commuting limits}\label{sec:QuasiBIC}

 Interpreting the concept of quasi-BIC requires understanding it as a state within the linear space of modes. 
 To see this state brought in by a small $\theta$, we examine the output state at $z_{\text{min}}$ when the propagating plane wave $W_1$ lands on the device at $z_{max}$
\begin{align}\label{eq:quasi BIC}  \text{Out}_{\text{min}}=S_{W_1W_1}\,W_1+S_{W_2W_1}\,W_2+S_{W_4W_1}\,W_4\;.
\end{align}

We observe that for 
$\theta=0$, 
 $S_{W_1W_1}\neq0$ but 
$S_{W_2W_1}=S_{W_4W_1}=0$, due to the presence of 
$\theta$ as a standalone factor in equations (\ref{Sw2w1}) and (\ref{Sw4w1}). 
Additionally, at 
$\theta=0$ and  $\Delta_a=0$, we notice  L'Hôpital's rule, resulting in $S_{W_1W_1}=1/A_{11}$. The output state becomes a pure propagating wave
\begin{align}\label{eq:quasi BIC theta Zero}  \text{Out}_{\text{min}}=(A_{11})^{-1}\,W_1
\end{align}
bearing no visible trace of the evanescent waves $W_2$ and $W_4$ that characterize the BICs.

The existence of the quasi-BICs is demonstrated by the ability to eliminate the propagating mode $W_1$ at $z_{min}$ when $S_{W_1W_1}=0$.
The elimination\textemdash occurring at two frequencies 
 $\Delta_a=\text{zero}_{1}$
  or 
$\text{zero}_{2}$
  ((\ref{eq:Zero1For2BICs}) or (\ref{eq:Zero2For2BICs})), if we stay in the frequency region around 
$\Delta_a$\textemdash results in the output field being a superposition of only the evanescent modes 
$W_2$ and  $W_4$.
This quasi-BIC state  warrants an exploration into the limit $\theta \rightarrow 0$. 
The key point with this limit is that it is constrained by $\Delta_a=\text{zero}_{1}=-\frac{29}{18} \theta^2 - \left(-\frac{\theta^2}{116197}\right)$ (taking $\text{zero}_{1}$ as an example), which represents a selected path in the $(\theta,\Delta_a)\text{-plane}$. At the heart of it is that the singularity introduced into the transfer matrix makes the limit of the scattering matrix  path-dependent in this plane.
A practical numerical method that we implemented using 3D electromagnetic simulation software \cite{studios2022cst} is to observe this path-dependent effect by tracking the limit behavior of the transmission scattering element $S_{W_1W_1}=S_{11}$ towards point $(\theta=0,\Delta_a=0)$ along the two paths depicted in Fig.\ref{fig:NonCommutingLimitsWithPaths}.

The constant value of $S_{W_1W_1}=0$ along the parabola ensures  a limit of zero, meaning complete opacity.
However, if we first set $\theta=0$  and then follow the vertical frequency axis until $\Delta_a$ reaches $0$, we end up with $S_{W_1W_1}=\frac{15}{17}+ \frac{9}{23} i$.
This value indicates that the device is almost transparent. 

The typical switching effect from opacity to transparency in the presence of a BIC can be summarized as 
\begin{align}\label{noncommuting_limits_S11}
   0= \lim_{\text{parabola}}S_{W_1W_1}\neq\\\nonumber
   \lim_{\Delta_a\rightarrow 0}\lim_{\theta\rightarrow 0}S_{W_1W_1}=\frac{15}{17}+ \frac{9}{23} i\;.
\end{align}
This noncommuting limit is readily visible in our numerical simulation Fig.\ref{fig:NonCommutingLimitsWithPaths} as well as in a variety of  experiments.

\section{Analysis  of the resonance and its accompanying BIC at the symmetric pattern frequency $\Omega_s$}\label{sec:ResonanceBIC}

\begin{widetext}
  \begin{minipage}{1\linewidth} 
    \begin{equation}\label{eq:TW1W3W5}
T^{-,-}_{W_1W_3W_5}=
\begin{pmatrix}
\scriptstyle{a_{11} + \Delta_s \beta_{11} + \theta^2 \mu_{11}} & \scriptstyle{-\sqrt{2}(a_{12} + \Delta_s \beta_{12} + \theta^2 \mu_{12} + \varphi^2 \tau_{12})} & \scriptstyle{-\sqrt{2}\varphi \nu_{14}} \\
\scriptstyle{-\sqrt{2}(a_{21} + \Delta_s \beta_{21} + \theta^2 \mu_{21} + \varphi^2 \tau_{21})} & \scriptstyle{\Delta_s \beta_{22} - \Delta_s \beta_{32} + \theta^2 \mu_{22} - \theta^2 \mu_{32}} & \scriptstyle{0} \\
\scriptstyle{-\sqrt{2}\varphi \nu_{41}} & \scriptstyle{0} & \scriptstyle{\Delta_s \beta_{22} - \Delta_s \beta_{32} + \theta^2 \mu_{44} - \theta^2 \mu_{54}}
\end{pmatrix}.
\end{equation}
  \end{minipage}
\end{widetext}

The resonance and its accompanying BIC are located around the frequency $\Omega_s$ that corresponds to the symmetric pattern of the transfer matrix.
The mathematical concept that comes to mind when discussing resonance is usually one pole close to the real axis in a complex plane. However, the scattering matrix elements developed below show that actually we are dealing with a Fano resonance, for which a pair of a pole and a zero is called into play.

To study the  resonance, $W_3$, and its associated  BIC, $W_5$ we need to work on the 3 by 3 matrix that shows the coupling between the incoming mode $W_1$ and the pair $W_3$, $W_5$. 

Around $\Omega_s$, the 3 by 3 matrix based on the subspace spanned by $(W_1,W_3,W_5)$ looks like in (\ref{eq:TW1W3W5}). 
Note that, contrary to (\ref{MatrixTw1w2w4}), $\varphi$ is part of the matrix $T^{-,-}_{W_1W_3W_5}$. The meaning of the symbols, as in (\ref{MatrixTw1w2w4}), comes from the series expansion (\ref{eq:Tseries}).
There are special relations $\nu_{14}=a_{12}$ and $\nu_{41}=a_{21}$ 
that will be used in the following computations. However, we did not use them in (\ref{eq:TW1W3W5}) to make clear the origin of each term in relation to the series expansion (\ref{eq:Tseries}).

We already see that, for $\varphi=0$, this matrix becomes block-diagonal, with $W_5$ decoupled from $W_1$ and $W_3$. In other words, to transform $W_5$ from a BIC into a quasi-BIC we need to switch $\varphi$ away from zero.
Later, we will see the same phenomenon and the need for a non-zero $\theta$ as well, through the zeros and poles of scattering matrix elements.

As in the two BICs case, here also we start from the two roots that correspond to $W_3$ and $W_5$, from the diagonal elements of (\ref{eq:TW1W3W5}). For the BIC the root is
\begin{equation}\label{eq:rootW5}
   \text{root}_{W5}=  -\theta^2 \frac{\mu_{44} - \mu_{54}}{\beta_{22} - \beta_{32}}=\frac{1}{45}\theta^2. 
\end{equation}

Resolving the zeros and poles related to the resonance entails a procedure distinct from merely solving for the root of the diagonal element $(W_3,W_3)$, which results in the following root being placed on the real frequency axis

\begin{equation}
\text{root}_{W3\text{real}}=-\theta^2 \frac{\mu_{22} - \mu_{32}}{\beta_{22} - \beta_{32}}=\left(\frac{1}{45} - \frac{2}{3}\right) \theta^2.
\end{equation}

This distinction arises from our earlier discussion of the resonance  at $\Omega_s$ and $(\theta=0,\varphi=0)$, which involved the 2x2 submatrix of the matrix (\ref{MatrixResonance}) that couples  $W_1$ and $W_3$. 
Here, for the quasi-BIC regime $(\theta\neq0,\varphi\neq0)$, the 2 by 2 matrix that couples $W_1$ and $W_3$ in $T^{-,-}_{W_1W_3W_5}$ is
\begin{equation}\label{matrixW1W3}
    \begin{pmatrix}
a_{11} & -\sqrt{2}a_{12} \\
-\sqrt{2}a_{21} & \Delta_s \beta_{22} - \Delta_s \beta_{32} + \theta^2 \mu_{22} - \theta^2 \mu_{32}
\end{pmatrix}.
\end{equation}
Notice that we neglect the small numbers $\Delta_s$, $\theta$ and $\varphi$ with respect to $a_{11},a_{12}$ and $a_{21}$. This time, from $\text{det}(\text{matrix} (\ref{matrixW1W3}))=0$, by keeping the zero-order approximation in $\theta$, the root is a complex number
\begin{equation}
   \text{root}_{W3\text{complex}}= \frac{2\, a_{12}\, a_{21}}{a_{11} (\beta_{22} - \beta_{32})}=\frac{1}{16925} - \frac{i}{4429}
\end{equation} given that the numbers $a_{mn},\; m,n=1,2$
are complex, as they relate to the incoming propagative BF mode $(0,0)$. This complex root is close to  $\Delta_s=0$, which is the frequency (\ref{eq:rootW5}) at which the quasi-BIC goes into a BIC.

From these three roots we construct the zeros and poles of the scattering matrix.
Both poles can be further corrected through a series approximation. The correction to $\text{pole}_{W_3}$ is not significant for small $\theta$ and $\varphi$, so
\begin{equation}
    \text{pole}_{W3}= \text{root}_{W3\text{complex}}.
\end{equation}

However, the correction to $\text{pole } W_5$ is significant and meaningful. It shows the importance of the angle $\varphi$ so that the BIC becomes visible in the transmission, $S_{11}$,
\begin{align}
    \text{pole}_{W5}= -&\theta^2 \frac{\mu_{44} - \mu_{54}}{\beta_{22} - \beta_{32}}+\\\nonumber
   & \theta^2 \varphi^2\frac{ -\mu_{2,2} + \mu_{3,2} + \mu_{4,4} - \mu_{5,4}}{\beta_{2,2} - \beta_{3,2}}.
\end{align}
The analytic formula reveals a relationship between the real roots and the real  $\text{pole}_{W5}$
\begin{equation}
    \text{pole}_{W5}=\text{root}_{W5} + \varphi^2 (\text{root}_{W3\text{real}} - \text{root}_{W5})
\end{equation}
which is preserved by the rational numbers we are using. With this at hand, we thus have
\begin{equation}
    \text{pole}_{W5}= \frac{1}{45}\theta^2 -\frac{2}{3}\theta^2 \varphi^2
\end{equation}
and from the cofactors, two  zeros  that are close to $\Delta_s=0$ for $\theta\approx 0$
\begin{equation}
    \text{zero}_{W3}=\text{root}_{W5}=\frac{1}{45} \theta^2
\end{equation}
and
\begin{equation}
\text{zero}_{W5}=\text{root}_{W3\text{real}}=\left(\frac{1}{45} - \frac{2}{3}\right) \theta^2\;.
\end{equation}
\begin{widetext}
  \begin{minipage}{0.8\linewidth} 
    
The scattering matrix elements that couple $W_1$ at $z_{max}$ into $W_3$ and $W_5$ at $z_{min}$ are

\begin{align}
  S_{W_3W_1}=  \frac{{a_{21} \sqrt{2}}}{{-2 a_{21} \beta_{12} - 2 a_{12} \beta_{21} + a_{11} (\beta_{22} - \beta_{32})}} \frac{{\Delta_s - \text{zero}_{W3}}}{{(\Delta_s - \text{pole}_{W3}) (\Delta_s - \text{pole}_{W5})}}
\end{align}
\begin{align}
   S_{W_5W_1}= \varphi\;\frac{{\nu_{41} \sqrt{2}}}{{-2 a_{21} \beta_{12} - 2 a_{12} \beta_{21} + a_{11} (\beta_{22} - \beta_{32})}} \frac{{ \Delta_s - \text{zero}_{W5}}}{{(\Delta_s - \text{pole}_{W3}) (\Delta_s - \text{pole}_{W5})}}.
\end{align}
From one of the special relations mentioned earlier, $\nu_{41}=a_{21}$, we see that the constant factors above are equal to each other. The transmission of the incoming propagative mode is
\begin{align}
    S_{W_1W_1}= \frac{{\beta_{22} - \beta_{32}}}{{-2 a_{21} \beta_{12} - 2 a_{12} \beta_{21} + a_{11} (\beta_{22} - \beta_{32})}} \frac{{(\Delta_s - \text{zero}_{W5}) (\Delta_s - \text{zero}_{W3})}}{{(\Delta_s - \text{pole}_{W3}) (\Delta_s - \text{pole}_{W5})}}.
\end{align}
Numerical values help to distinguish the distinct behavior of the resonance versus the BIC
\begin{align}\label{SW3W1}
S_{W_3W_1}=\left(\frac{1}{637} + \frac{i}{2258}
\right) \frac{{\Delta_s - \frac{1}{45} \theta^2}}{{(\Delta_s - \left(\frac{1}{16925} - \frac{i}{4429}\right)) \left(\Delta_s - \frac{1}{45} \theta^2 + \frac{2}{3} \theta^2 \varphi^2-\delta\right)}}
\end{align}
\begin{align}\label{SW5W1}
   S_{W_5W_1}=\varphi\;\left(\frac{1}{637} + \frac{i}{2258}
\right) \frac{ \Delta_s - \left(\frac{1}{45} - \frac{2}{3}\right) \theta^2}{(\Delta_s - \left(\frac{1}{16925} - \frac{i}{4429}\right)) (\Delta_s - \frac{1}{45} \theta^2 + \frac{2}{3} \theta^2 \varphi^2-\delta)}
\end{align}
\begin{align}\label{eq:S11ResBIC}
 S_{W_1W_1}=  \left(\frac{23}{25} + \frac{34i}{101}
\right) \underbrace{\frac{\Delta_s - \left(\frac{1}{45} - \frac{2}{3}\right) \theta^2}{\Delta_s - \left(\frac{1}{16925} - \frac{i}{4429}\right) }\;}_{\text{resonance}}\;\underbrace{
\frac{\Delta_s - \frac{1}{45} \theta^2}{ \Delta_s - \frac{1}{45} \theta^2 + \frac{2}{3} \theta^2 \varphi^2-\delta}}_{\text{quasi-BIC}}.
\end{align}
  \end{minipage}
\end{widetext}

A very small additive correction, $\delta$, to the pole of the quasi-BIC arises from higher-order approximations that have been neglected so far.
We mention this correction because the displayed pole is not a complex number, as it should be due to the coupling of the evanescent waves to the propagative BF mode 
$(0,0)$. The numerical value of this correction is
\begin{equation}\label{eq:delta_correction}
\begin{split}
   \delta =& \theta^2 \varphi^2 \\
   &\left(- \left( 370 + 1827 i \right) \theta^2 +\left( \frac{1}{1046639} + \frac{i}{282269} \right) \varphi^2 \right).
\end{split}
\end{equation}
This correction is fourth order in $\theta$ and $\varphi$, with the term $\theta^4\varphi^2$ dominating the one corresponding to $\theta^2\varphi^4$.
The symbolic formula for the correction $\delta$ is placed in Appendix \ref{Appendix:delta}.



\begin{figure}
    \centering
    \includegraphics[width=1\linewidth]{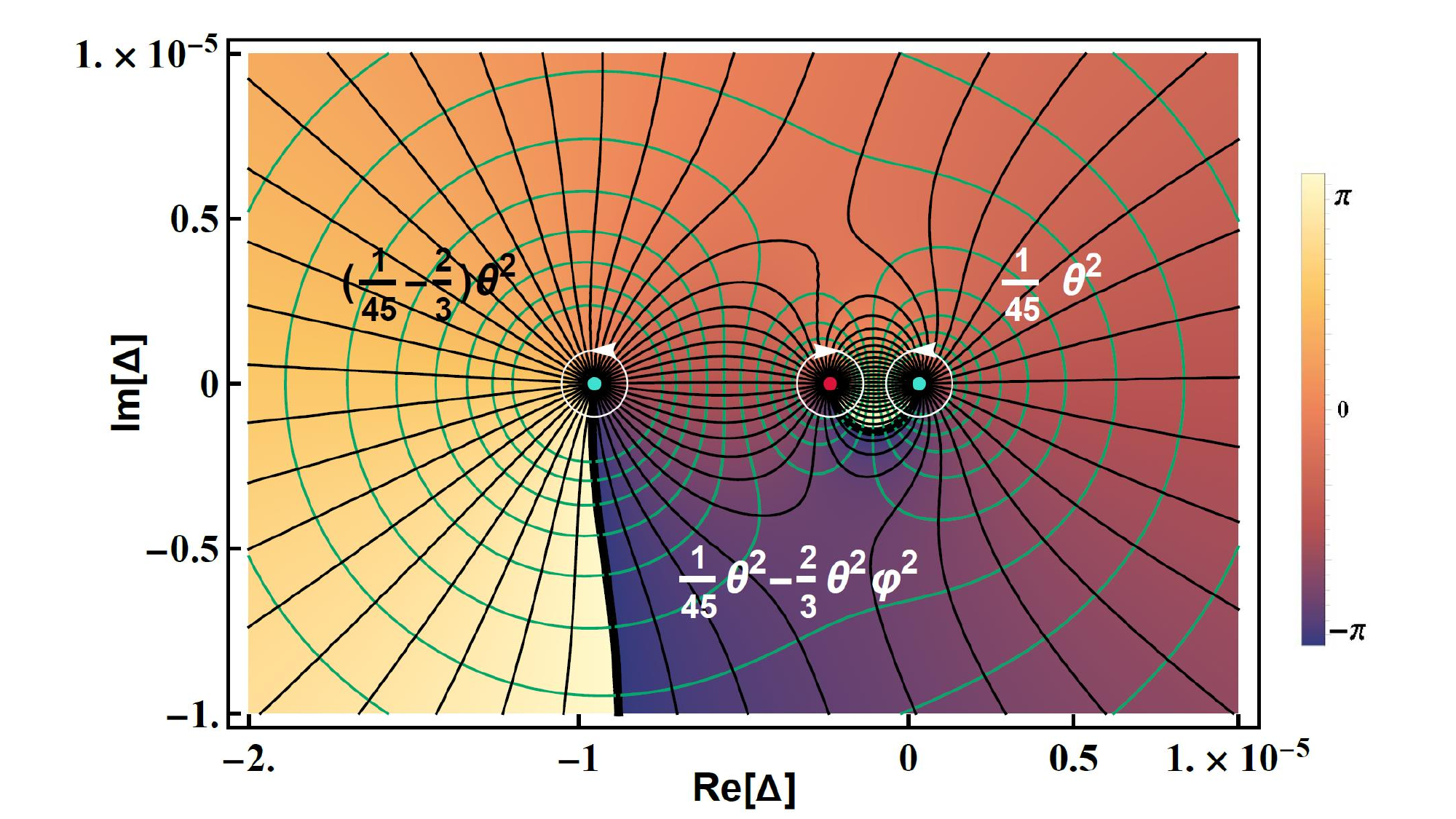}
\captionsetup{justification=raggedright,singlelinecheck=false}
    \caption{The electrostatic model for resonance and its quasi-BIC. The angle-dependent zeros and pole of 
$S_{W_1W_1}$ from (\ref{eq:S11ResBIC}) were selected to be displayed within the plotted region, while the second pole lies outside this area. The plot shows contour lines for the complex electrostatic potential $\ln(S_{W_1W_1})$.
    The real part of the complex potential is in emerald green, whereas the imaginary part in black. At each three singular point a charged line, perpendicular to the plane, is located.  The zeros, marked with green disks have a  charge of +1, whereas the pole, marked red, has a  charge of -1. 
 The position on the real axis of these three singularities are written in terms of the angles $\theta$ and $\varphi$.}
    \label{fig:BICcomplexPlane}
\end{figure}



 
Studying these matrix elements provides the insights necessary to refine the initially preliminary concept of BIC at  $\Omega_s$ into a more specific and detailed meaning.

This meaning connects the null subspace of the transfer matrix perspective with the positioning of zeros and poles in the complex frequency plane.

For $\varphi=0$, we observe that the numerator simplifies in $S_{W3W1}$ (\ref{SW3W1}) even when  $\theta \neq 0$, resulting in the elimination of the BIC from this scattering matrix element. However, in $S_{W5W1}$ (\ref{SW5W1}), such simplification does not occur; instead, the presence of $\varphi$ in the numerator renders the entire scattering matrix element null at $\varphi=0$. We conclude that the BIC  can transition into a quasi BIC only in the presence of nonzero $\theta$ and $\varphi$. 
 The dependence on $\varphi$ was also inferred directly from the transfer matrix (\ref{eq:TW1W3W5}), which turns block-diagonal for $\varphi=0$.

The signature of the quasi-BIC in $S_{W1W1}$ springs from the term $(\Delta_s - \frac{1}{45} \theta^2)/( \Delta_s - \frac{1}{45} \theta^2 + \frac{2}{3} \theta^2 \varphi^2-\delta)$, which is visible in transmission as a deep, narrow notch as long as $\theta \approx 0$  and $\varphi\approx 0$, but not zero, as shown in panel (a) of Fig.\ref{fig:All3BICSandResonance}.

Significantly, the same term simplifies to 1 when $\theta=0$, yet the factor $\Delta_s - \left(\frac{1}{45} - \frac{2}{3}\right) \theta^2$ persists in the numerator due to the presence of the complex pole in the denominator. The occurrence of  this factor in the numerator of $S_{W1W1}$ is reflected in transmission  dropping to zero observed in both theoretical models and numerical simulations as shown in panel (c) of Fig.\ref{fig:All3BICSandResonance}.  The identical phenomenon is absent in the case of two BICs around $\Omega_a$, as the quadratic $\Delta_a$  in the numerator simplifies with a quadratic $\Delta_a$  in the denominator, (\ref{Sw1w1}).
The complex pole at $\theta=0, \varphi=0$ is distinctly observable in  $S_{W3W1}$, 
given that is is very close to the real frequency axis.
The disparity in the visibility of the complex pole  in  $S_{W3W1}$ but not in  $S_{W5W1}$ at $\theta=0, \varphi=0$, further strengthens our association of the term resonance with the mode $W_3$, stemmed from the transfer matrix in Section \ref{sec:THETA_Zero}.

We turn to the discussion of the Q-factor, as it emerges from the scattering matrix elements.
As an example for this topic, consider the quasi-BIC linear fractional function from (\ref{eq:S11ResBIC}).

When the zero $z$ is close to the pole $p$ in  $(\Omega-z)/(\Omega-p)$, they can create sharp peaks or troughs in the frequency response curve. 
In line with the chosen example, the zero 
$z$ is set to be a real number, while the pole 
$p$ is allowed to be a complex number.
Denote $\Omega_\text{max}$ and $\Omega_\text{min}$ the frequencies at which $|\Omega-z|^2 /|\Omega-p|^2$
attains a maximum and a minimum on the real frequency axis, respectively. The narrowness of the notches observed in the transmission due to a quasi-BIC is measured via a Q-factor
The formulas for the Q factor depend on the type of system and its implementation, but generally, they all relate the bandwidth to the center frequency.
Define the Q-factor in terms of the frequencies $\Omega_\text{max}$ and $\Omega_\text{min}$ as
\begin{align}\label{DefinitionQ}
Q=\frac{\text{central frequency}}{\text{bandwidth}} :=
\frac{2^{-1}(\Omega_\text{max}+\Omega_\text{min})}{\left|\Omega_\text{max}-\Omega_\text{min}\right|}.  
\end{align}
We find that, for the zero placed on the real axis, 
\begin{align}
Q=\frac{\left|z^2-\left|p\right|^2\right|}{\left|z-p\right|^2}.
\end{align}
Applying this formula to the quasi-BIC from (\ref{eq:S11ResBIC}), where 
\begin{align}
z&=\Omega_s+\frac{1}{45} \theta^2\\
p&=\Omega_s+\frac{1}{45} \theta^2- \frac{2}{3} \theta^2 \varphi^2+\delta
\end{align}and neglecting 
$\delta$ and a small term relative to 
$\Omega_s$, the Q-factor can be expressed as 
\begin{equation}\label{Qfactor}
    Q=3 \frac{\Omega_s}{ \theta^2 \varphi^2}.
\end{equation}

 The Q-factor goes to infinity diverging as a second power in the product of the angles.

The transition from a quasi-BIC to a BIC can be understood through an analogy with an electrostatic model Fig.\ref{fig:BICcomplexPlane}, as shown in \cite{grigoriev2013singular}.

As $\varphi$ goes to zero, the rightmost charges +1 and -1 come closer to each other, annihilating at $\varphi=0$. This annihilation marks the transition of the quasi-BIC into a BIC. The zero from the left is not annihilated even if both angles become zero. 

This is responsible for the resonance dip in the transmission $S_{W_1W_1}$, as shown in panel (c) of Fig.\ref{fig:All3BICSandResonance} and described in equation (\ref{eq:S11ResBIC}).

The correction  (\ref{eq:delta_correction}) to the quasi-BIC's pole is not shown in Fig.\ref{fig:BICcomplexPlane}, being too small to be visible. The parameter $\Delta=\Delta_s$ was scaled, and $\theta$ and $\varphi$ were chosen to facilitate the visualization of the singular points, given that these adjustments  does not influence the values of the charges.

\section{Impact of higher Bloch-Floquet modes and comparison to numerical simulations}\label{sec:higher modes}

\begin{figure}
    \centering
    \includegraphics[width=1\linewidth]{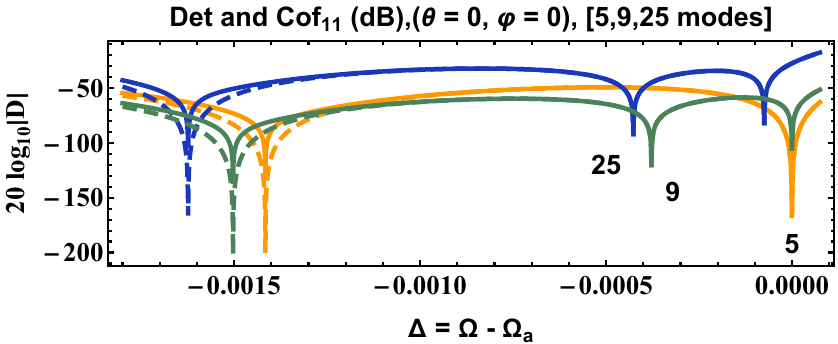}
\captionsetup{justification=raggedright,singlelinecheck=false}
    \caption{ On the vertical axis, on a logarithmic scale, $D$ represents a pair formed by the determinant of $T^{-,-}$, plotted as a continuous line, and the cofactor of its $T_{11}^{-,-}$  matrix element, plotted as a dashed line. Three pairs of curves are represented, corresponding to 5, 9, and 25 modes. The curves belonging to the same pair superimpose, showing that they indeed share common zeros at $\theta=0$ and $\varphi=0$. The region on the right side of the frequency axis belongs to the antisymmetric pattern where the two BICs are located. For the 5-dimensional case, at $\theta=0$ and $\varphi=0$, these two BICs coincide, as Fig. 4 also shows. They become distinct as quasi-BICs. However, increasing the number of modes beyond 5 causes the BICs to be located, at $\theta=0$ and $\varphi=0$, at distinct frequencies, although close to each other. The region on the left side of the frequency axis belongs to the symmetric pattern where the resonance and its BIC are located.
     The BIC that accompanies the resonance is  visible for $\varphi=0$ through the rate of decrease being larger for the cofactor than for the determinant around the roots. }
    \label{fig:DetCoff11ManyModes}
\end{figure}

Once the behavior of the device is understood in the 5-dimensional subspace of modes, it becomes necessary to see if the findings survive in a larger dimensional subspace. In Fig.\ref{fig:DetCoff11ManyModes}, the effect of enlarging the subspace dimension is visible.
At  $\theta=0$ and $\varphi=0$ , the BICs are not visible in $S_{11}$. However, the advantage of our approach is that we can plot the numerator and the denominator of $S_{11}$ separately, so the poles and zeros appear with full visibility, even if they are equal. Note that Fig.\ref{fig:DetCoff11ManyModes} uses a logarithmic scale on the vertical axis, unlike the linear scale in Fig.\ref{fig:Det5by5}. The logarithmic scale helps the zeros and poles appear as thin notches, making it easier to compare their positions as the number of modes increases.

The right side of the frequency axis, in the vicinity of $\Omega=\Omega_a$, corresponds to the antisymmetric pattern. Here, for 5 modes, both the determinant and the cofactor have a real double root in $\Delta_a$, as we discuss in relation to Fig. 4. This double root property does not survive in the presence of a larger number of modes. The double root splits into two simple real roots, visible for 9 and 25 modes. In logarithmic scale, the distinction between a double root and a simple root is that the notch of a double root drops faster. Two BICs are still present in the frequency region of the antisymmetric pattern, but they are positioned at different frequencies, which become stable by 25 modes  at $\Omega_{a1}=0.05126348554915704$ and $\Omega_{a2}=0.05088560198916318$.

The left side of the frequency region in Fig.\ref{fig:DetCoff11ManyModes} is the location of the resonance and its accompanying BIC. The cofactor has a real double root in 5 dimensions, a property that is maintained as the number of modes increases. The determinant has a simple real root, and so it drops more slowly than the cofactor's notch. This root shifts to the lower frequency as the number of modes increases, eventually reaching the value $\Omega_s=0.04976062158072677$ by 25 modes Fig.\ref{fig:DetCoff11ManyModes}. The accompanying BIC is not separately visible because, for $\theta=0$ and $\varphi=0$, it becomes the second of the double roots of the cofactor.

\begin{figure}[h]
    \centering
    \begin{subfigure}[b]{0.48\textwidth}
        \centering
        \includegraphics[width=1.0\textwidth]{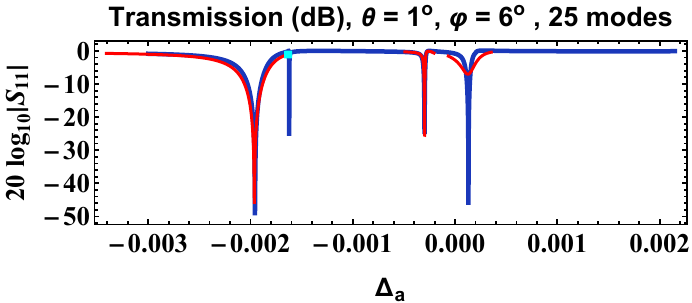}
\captionsetup{justification=raggedright,singlelinecheck=false}
        \caption{The theoretical results are plotted in blue with respect to $\Delta_a=\Omega-\Omega_a$. The numerical simulations are plotted in red, except for the quasi-BIC at resonance, which is shown in cyan and is very narrow, making it difficult to resolve in the numerical simulations. All four dips in transmission are present for $\theta\neq0$ and $\varphi\neq0$. Their order, from left to right, is as follows: resonance, quasi-BIC of the resonance, $\text{quasi-BIC}_{1a}$, and  $\text{quasi-BIC}_{2a}$ } 
        \label{fig:fig1}
    \end{subfigure}
    \vfill
    \begin{subfigure}[b]{0.48\textwidth}
        \centering
        \includegraphics[width=1.0\textwidth]{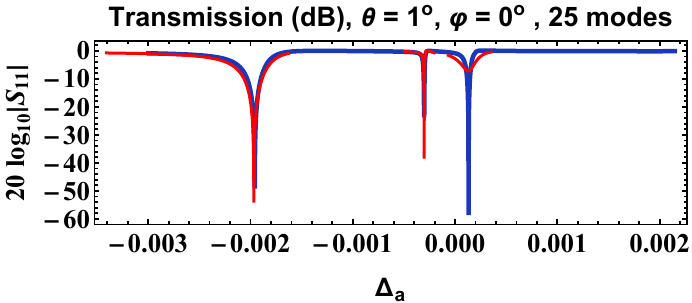}
\captionsetup{justification=raggedright,singlelinecheck=false}
        \caption{The quasi-BIC that accompanies the resonance becomes a BIC for $\varphi=0$, and is thus absent in transmission. The two quasi-BICs associated with the antisymmetric pattern are not affected by switching off $\varphi$.  }
        \label{fig:fig2}
    \end{subfigure}
    \vfill
 \begin{subfigure}[b]{0.48\textwidth}
        \centering
        \includegraphics[width=1.0\textwidth]{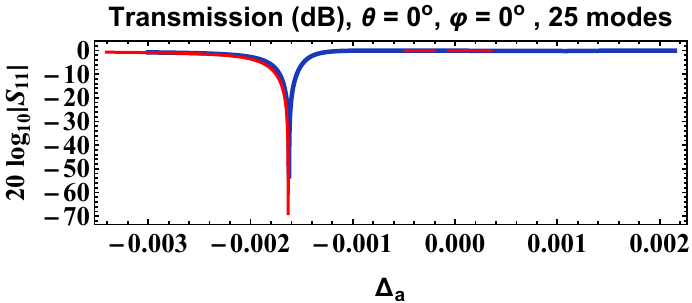}
\captionsetup{justification=raggedright,singlelinecheck=false} 
        \caption{For $\theta=0$ and $\varphi=0$, all three quasi-BICs become three distinct BICs. The resonance remains visible.}
        \label{fig:fig3}
    \end{subfigure}
\captionsetup{justification=raggedright,singlelinecheck=false}   
    \caption{The quasi-BICs convert into BICs as the angles are switched off. This property, designed using 5 modes, survives in the presence of a large number of Bloch-Floquet modes and is confirmed by numerical simulations. }
    \label{fig:All3BICSandResonance}
\end{figure}

The three panels of Fig.\ref{fig:All3BICSandResonance}, all obtained using 25 modes, show that the properties of the BICs in transmission, as derived with 5 modes, survive the impact of the large number of BF modes. To prevent any confusion when discussing the panels of this figure, it is crucial to clearly distinguish between a quasi-BIC and a BIC, even though the frequency labels do not include the 'quasi' qualifier.

Panel (a) shows the theoretical results plotted in blue. On the left, the two dips in blue correspond to the resonance  $f_{\text{Resonance}}=2352.44 \text{ GHz}$ and its accompanying quasi-BIC $f_{\text{BIC of Resonance}}=2368.66 \text{ GHz}$, both associated with the symmetric pattern of the transfer matrix. On the right, for the antisymmetric pattern, also in blue, the two quasi-BICs appear at $f_{\text{BIC}_{1a}}=2431.64 \text{ GHz}$ and $f_{\text{BIC}_{2a}}=2452.16 \text{ GHz}$.

The numerical simulations are plotted in red, except for the quasi-BIC associated with the resonance, which is plotted in cyan. All numerical simulations are systematically shifted to slightly higher frequencies. To superpose them on the theoretical dips, a downward shift of 20 GHz was needed for $\text{quasi-BIC}_{1a}$, whereas for the rest, it was 17 GHz, resulting in an error of about $0.8\%$ in this frequency band.

The numerical simulation dedicated to the quasi-BIC associated with the resonance was difficult to perform because its dip covers a range of less than 10 MHz at a frequency of $f_{\text{BIC of Resonance}}=2368.66 \text{ GHz}$. This presented a clear advantage for the theory, as setting up a numerical simulation without it would have been highly impractical. The numerical simulation plotted in cyan appears more like a dot than a dip because it lacks the vertical range seen in the corresponding theoretical plot, which is shown in blue.

Panels (b) and (c) confirm that the quasi-BICs convert into BICs when one or both of the angles are set to zero, even in the presence of a large number of BF modes.

Besides considering the influence of the number of BF modes, we are also interested in the impact of large deviations of the angles from zero, which push the BF evanescent mode $(-1,0)$ close to becoming propagative. The discrepancy between the theoretical and numerical dips in transmission seen in panel (a) of Fig.\ref{fig:All3BICSandResonance} for $\text{quasi-BIC}_{2a}$ reduces as the angles approach zero.


\section{Conclusion and Discussions}\label{sec:ConclusionsDiscussions}

In what follows, we will discuss the main topics of this paper. 

The first one is the concept of BIC and we suggest that it is not trivial to define it in the general sense.
The concept of BIC is often illustrated through a figure representing energy levels classified as extended states (continuum), resonance (leaky modes), BICs, and regular bound states (discrete levels), like the one in Figure 1 of reference \cite{hsu2016bound}. This type of figure is used as a definitional starting point in many papers, providing a visual representation of the BIC.
Unlike the visual-based definition, the spectrum of a scattering problem is mathematically described from the functional analysis perspective as in \cite{Simon1978overview}, which indicates that the spectrum is composed of three distinct parts:  pure point, absolutely continuous  and singular continuous. In between, studies like \cite{evans1998trapped}, \cite{mciver2001embedded},\cite{evans1994existence}
\cite{mciver1998construction}, and \cite{dhia2018trapped} define a specific mathematical-physics model and an eigenvalue problem associated with it. This discussion highlights the variety of approaches employed to introduce the BIC concept. Importantly, the effectiveness of a chosen framework hinges on the specific goals of the research project.

Regardless of the initial definition used, it turns out that  BICs share a set of common properties. 

Localization: BICs are spatially localized  even though they coexist with a continuous spectrum of radiating waves that can carry energy away.

Real eigenvalue: BICs have a real-valued frequency eigenvalue.

Non-interacting: BICs do not interact with the states of the continuum. This means they cannot decay or be excited by external waves.

Infinite Q factor: The Q factor is a measure of the resonance quality of a state.  In an ideal BIC (without any absorption), the Q factor would be infinite, signifying a perfectly sharp resonance.

It is important to note that these properties of BICs are interrelated. While the list doesn't imply a causal sequence (one property doesn't necessarily lead to another), they all contribute to the unique behavior of BICs. 

In this paper, only the first BF mode, $(0,0)$, out of the five selected for the 5x5 matrix (\ref{TmatrixThetaZero}) is propagative at $\theta=0$ and $\varphi=0$. The other four modes are evanescent in the $z$-direction, and so any linear combination of these four evanescent modes decays along the $z$-direction. This attenuation is what we mean by being spatially localized under the localization property.

The BIC state will emerge as a linear combination of these evanescent modes. These evanescent modes coexist with a continuous spectrum of radiating waves capable of carrying energy away, with the BF $(0,0)$ mode being the sole radiative one, in our case. This coexistence is reflected in the transfer matrix elements from the first row and first column in (\ref{TmatrixThetaZero}), which capture, in terms of complex numbers, the couplings between the evanescent modes and the propagative mode.

To implement the real eigenvalue property, we enforce a non-empty null space for the transfer matrix, requiring the presence of a zero eigenvalue (see (\ref{eq:DetCof}) and (\ref{eq:CofZero})). By doing this, we ensure that the zeros of the scattering matrix elements are  on the real axis of the complex frequency plane for $\theta=0$ and $\varphi=0$.
 
We dedicated Section \ref{sec:QuasiBIC} to the non-interacting property of the BIC. This property is inherently connected to the concept of quasi-BIC. 
The BIC state can be seen as the limit of a sequence of quasi-BIC states, which in our case are parameterized by the angles, although other parameters could also be used (e.g., geometrical or material).
When the sequence of quasi-BIC modes reaches the BIC limit, the transmission scattering matrix of the BF mode $(0,0)$ shows no trace of the evanescent waves, indicating that the interaction of the BIC with the continuum state is nullified (\ref{eq:quasi BIC theta Zero}). Furthermore, the concept of non-interaction is actually more nuanced, captured in the non-commuting limit (\ref{noncommuting_limits_S11}) and in Fig.\ref{fig:NonCommutingLimitsWithPaths}.

Finally, the infinite Q-factor property is linked to the Fano resonance, described here as a linear fractional function of the frequency. The Q-factor, as defined in (\ref{DefinitionQ}), is inversely proportional to the distance between the zero and the pole of this function. For a quasi-BIC, this distance varies proportionally with powers of  $\theta$ and $\varphi$. As quasi-BIC approaches BIC, this distance tends to zero, resulting in an infinite Q-factor (\ref{Qfactor}).

We now turn our attention to the methodology  employed in this paper, contrasting it with temporal coupled-mode theory (TCMT)\cite{joannopoulos2008molding}. TCMT adopts a fundamental approach, modeling the system as a collection of key components governed by established principles such as energy conservation, linearity, and weak coupling. The building blocks include localized and propagating modes. This approach leads to a general description applicable to a wide range of photonic devices. To facilitate quantitative analysis, this description is expressed in terms of many fitting parameters \cite{grigoriev2013optimization}. The need to reduce the number of fitting parameters brings effective Hamiltonians into play, as well as other phenomenological approaches.

For instance, in \cite{zhen2015spawning}, TCMT is used to generate an effective Hamiltonian, with matrix elements derived from experimental data by fitting reflectivity curves. Once the parameters of the effective Hamiltonian are obtained, its eigenvalues can be calculated. The close match between these calculations and experimental results validates the TCMT equations.

TCMT offers significant benefits: it possesses both predictive and explanatory power. On one hand, it allows predicting the device's response to various inputs; on the other hand, it provides insights into how different modes within the device interact. This dual capability is a defining characteristic of gray box models.

In contrast to TCMT, our model is analytical, relying on closed-form solutions derived from first principles. Similar to TCMT's weak coupling approximations, our solutions for Maxwell's equations on a lattice require that the device's thickness along the 
$z$-axis be small compared to the unit cell dimensions. 
The accuracy of this approach is validated by comparing it with numerical simulations presented both in this paper and in \cite{Lipan:24}.

Our method  is broadly applicable, extending beyond the specific device shown in Fig.\ref{fig:Device}. The equations connecting geometrical parameters and frequencies have clear, well-defined meanings, allowing for straightforward interpretation of the results. Additionally, our analytical solutions are computationally efficient, providing quick answers without requiring extensive resources, as demonstrated in the Section\ref{sec:higher modes} where 25 modes were used. The theory can capture very narrow bandwidth BICs on the frequency axis, guiding the subsequent search with numerical software. This is not always an easy task for a full-wave CAD software package, not to mention that searching for a BIC that manifests within a 10 MHz bandwidth, embedded somewhere in a frequency region of interest spanning several THz, is not trivial either. These precise analytical solutions can serve as benchmarks for validating more complex numerical or empirical models, testing their accuracy and reliability.

As for what we did not touch on in this paper but will address, it includes the study of BICs away from the $\Gamma$-point and for different devices, such as those with a non-square unit cell.
Walking on the surface from Fig.\ref{fig:SurfaceThetaZeroViewPoint2}, away from the SincSync point, expanding on the topological properties of the proposed devices in relation to the dispersion relations, and examining them from the BICs' classification point of view are other avenues to explore.


\section{Acknowledgment}
 We wish to thank Dr. Andrei Silaghi, with Continental Automotive Romania SRL, for facilitating the access to numerical simulation using CST microwave studio \cite{studios2022cst}.

\appendix

\begin{widetext}

\section{Symbolic formulas for the poles of the antisymmetric pattern}\label{AppendixPolesAntisymmetric}

\begin{align}
  \text{pole}_{1}&=\text{root}_{W2}-2\theta^2\,  \frac{
  2 \left( B_{1 4} B_{2 1} - A_{1 1} \mu_{2 4} \right) \mu_{4 2} + 
  B_{1 2} \left( 2 B_{4 1} \mu_{2 4} + B_{2 1} \left( \mu_{2 2} + \mu_{3 2} - \mu_{4 4} - \mu_{5 4} \right) \right)
}{
  \left( \beta_{2 2} + \beta_{3 2} \right) \left( 2 B_{1 2} B_{2 1} + 2 B_{1 4} B_{4 1} + A_{1 1} \left( \mu_{2 2} + \mu_{3 2} - \mu_{4 4} - \mu_{5 4} \right) \right)
}\\
\text{pole}_{2}&=\text{root}_{W4}+2\theta^2 \,\frac{
  2 \left( B_{1 2} B_{4 1} - A_{1 1} \mu_{4 2} \right) \mu_{2 4} + 
  B_{1 4} \left( 2 B_{2 1} \mu_{4 2} + B_{4 1} \left( -\mu_{2 2} - \mu_{3 2} + \mu_{4 4} + \mu_{5 4} \right) \right)
}{
  \left( \beta_{2 2} + \beta_{3 2} \right) \left( 2 B_{1 2} B_{2 1} + 2 B_{1 4} B_{4 1} + A_{1 1} \left( -\mu_{2 2} - \mu_{3 2} + \mu_{4 4} + \mu_{5 4} \right) \right)
}
\end{align}

\section{Symbolic formula for the correction to the quasi-BIC pole for the symmetric pattern}\label{Appendix:delta}

\begin{align}
&\delta=\theta^2 \varphi^2  
\frac{
\mu_{22} - \mu_{32} - \mu_{44} + \mu_{54}
}{
2 a_{12} a_{21} (\beta_{22} - \beta_{32})
}\\\nonumber
&\Bigg(
\theta^2 \bigg(
2 a_{21}( \mu_{12}+\frac{\beta_{12}}{\beta_{22} - \beta_{32}}(- \mu_{44} + \mu_{54})) +2 a_{12} (\mu_{21}+\frac{\beta_{21}}{\beta_{22} - \beta_{32}}(- \mu_{44} + \mu_{54}))\\\nonumber
&
- a_{11} (\mu_{22} - \mu_{32} - \mu_{44} + \mu_{54})
\bigg)+ 2 \varphi^2 (
a_{12} a_{21} + a_{21} \tau_{12} + a_{12} \tau_{21}
)
\Bigg)
\end{align}

\end{widetext}

\bibliography{BIC}

\end{document}